\documentclass[useAMS,usenatbib]{mn2e}
\usepackage[varg]{txfonts}
\usepackage{graphicx}
\usepackage{amssymb}
\usepackage{epstopdf}
\usepackage{gensymb}
\usepackage{subfig}
\usepackage{amstext}
\newcommand{\specialcell}[2][c]{
  \begin{tabular}[#1]{@{}c@{}}#2\end{tabular}}
\usepackage{url}

\title[BACHES - a compact \'{e}chelle spectrograph for radial velocity surveys with small telescopes]
{BACHES - a compact \'{e}chelle spectrograph for radial velocity surveys with small telescopes}
\author[S.~K.~ Koz\l owski, M. Konacki, M. Ratajczak, P. Sybilski, R.K. Paw\l aszek, and K.G. He\l miniak]
{
S. K. Koz\l owski$^{1}$\thanks{E-mail:
stan@ncac.torun.pl}, 
M. Konacki$^{1,2}$,
M. Ratajczak$^{1}$,
P. Sybilski$^{1}$,
R.K. Paw\l aszek$^{1}$,\newauthor
~and K.G. He\l miniak$^{3,1}$\\
$^{1}$Nicolaus Copernicus Astronomical Center, Toru\'{n}, Poland \\
$^{2}$Astronomical Observatory, Adam Mickiewicz University, S\l oneczna 36, 60-268 Pozna\'{n}, Poland\\
$^{3}$Subaru Telescope, National Astronomical Observatory of Japan, 650 North Aohoku Place, Hilo, HI 96720, USA \\
}

\begin{document}

\date{Accepted 2014 June 9.  Received 2014 June 9; in original form 2014 March 18.}

\pagerange{\pageref{firstpage}--\pageref{lastpage}} \pubyear{2014}

\maketitle

\label{firstpage}

\begin{abstract}
We evaluate a pre-production BACHES \'{e}chelle spectrograph in terms of its usefulness for radial velocity surveys of binary stars with small telescopes in a remote and autonomous. We use the Solaris-4 observatory located in CASLEO, Argentina, that is part of a global network of autonomous observatories as the test-bed for the instrument. The setup is designed in such a way that spectroscopy and photometry can be carried out using the same telescope without the need to mechanically modify the imaging train. We observe single spectroscopic standard stars as well as binary stars up to 9.75 mag. We present results of mechanical tests of the instruments and spectroscopic observations carried out between Nov 26th and Dec 8th 2013. We conclude that BACHES is a very compact and capable spectrograph well suited for remote and autonomous operation. Coupled to a 0.5-m telescope it is capable of obtaining spectra of 10 mag targets with a SNR of 20 for 30-minute exposures. This is a very good result considering the price and size of the instrument. It brings new possibilities to the scientific community and opens a whole new range of research opportunities available to new and existing observatories. 
\end{abstract}


\section{Introduction}
 
Photometric surveys are popular and valuable projects that aim at detecting exoplanets, variable stars, near Earth objects,  transient events and so forth. The main idea behind such surveys is to detect variability in brightness and/or position of a given object by comparing values of its parameters throughout the duration of the survey or with historical data. Photometric surveys can be devoted to a preselected group of targets that exhibit or may exhibit certain interesting properties that are worth observing or can simply rely on repetitive and continuous imaging of the entire visible sky (or selected region), e.g. the All Sky Automated Survey (ASAS) \citep{Pojmanski1997}, the Optical Gravitational Lensing Experiment (OGLE) \citep{Udalski1992}, the Wide Angle Search for Planets (SuperWASP) \citep{Street2003}, the Hungarian Automated Telescope (HAT) \citep{Bakos2013}. To get the complete set of stellar parameters for binary systems, however, multicolour photometry is not enough - spectroscopic follow-up observations are needed. Given the wide stream of photometry data available from ground- and space-based observations, it seems impossible to manually conduct spectroscopic follow-ups that would catch up with the photometric data flow. 

Spectroscopic observations present a higher level of complexity in terms of both hardware and software. Data acquisition is much more troublesome and often requires human intervention -- especially when using older instruments. Data reduction is more complicated than in the case of photometry, thus is more challenging to automate. Spectroscopy is also much more sensitive to all sorts of mechanical errors and instabilities due to the relatively longer exposure times and the use of slit or fibre that require precise positioning and tracking. Nevertheless, it seems inevitable to conduct automatic spectroscopic surveys in the era of remote and autonomous observatories. 

In a very comprehensive review, \cite{Torres2010} present results of several decades of photometric and spectroscopic observations that led to the determination of stellar parameters of 94 eclipsing binary systems with very high precision. Given that the periods of the binaries presented in the review are mostly below 10 days, it is straightforward to conclude that the effectiveness of the survey was limited by the availability of the instrument, observer and weather conditions. While the last factor is impossible to eliminate  (though distributed global networks of telescopes try to limit the influence of weather conditions on the observing efficiency), the way the instrument is used by the observer can be optimised. In the best case scenario, the role of the observer should be limited to selecting the targets and scheduling the observations. Target acquisition, instrument calibration and actual observing should be carried out automatically since these are all repetitive and well defined tasks. In practice, the world's largest telescopes equipped with spectrographs do not observe automatically. In case of large telescopes, the cost of the observer and operator are a relatively small share of the total operating costs of the instruments. The profit of automating the entire observing procedure with a complex and expensive instrument could be questionable. In case of smaller instruments, however, optimising efficiency will bring much profit and scientific payback. Good examples of autonomous or remotely operated spectrographs are the STELLar Activity (STELLA) telescopes \citep{Weber2012}, one of which is equipped with a spectrograph and operates autonomously, the Wide Field Spectrograph (WiFeS) \citep{Dopita2010} that is operated remotely on a regular basis and the Folded Low Order whYte-pupil Double-dispersed Spectrograph (FLOYDS) \citep{Sand2011} used by the Las Cumbres Observatory Global Telescope Network.

In this paper we describe results of tests conducted with a new compact \'{e}chelle spectrograph and a 0.5-m telescope. Our goal was to investigate if such setup is capable of producing satisfactory scientific results and if it is suitable for large scale spectroscopic surveys.  A somewhat similar approach is presented by \cite{Vanzi2012}, who describe PUCHEROS, a cost-effective $R\sim20,000$ fibre-fed \'{e}chelle spectrograph intended for use with small telescopes. \cite{Csak2014} present results of tests conducted with eShel, a commercially available fibre-fed  $R\sim11,000$ echelle spectrograph attached to 1-m and smaller telescopes. It should be noted that our setup is designed in such a way that spectroscopy and photometry can be carried out using the same telescope without the need to mechanically modify the imaging train.

In Sec. \ref{sec:SpectroscopicSurvey} we analyse the potential of an automated spectroscopic survey. In Sec. \ref{sec:BACHES} we describe the spectrograph, in Sec. \ref{sec:TestSite} we present the test site, including the instruments and the software setup. Test results are described in Sec. \ref{sec:TestResults}. We summarise our findings in Sec. \ref{sec:Summary}.

\section{Spectroscopic survey}
\label{sec:SpectroscopicSurvey}

\begin{figure}
\begin{center}
\includegraphics[bb=17 224 558 611,width=\columnwidth]{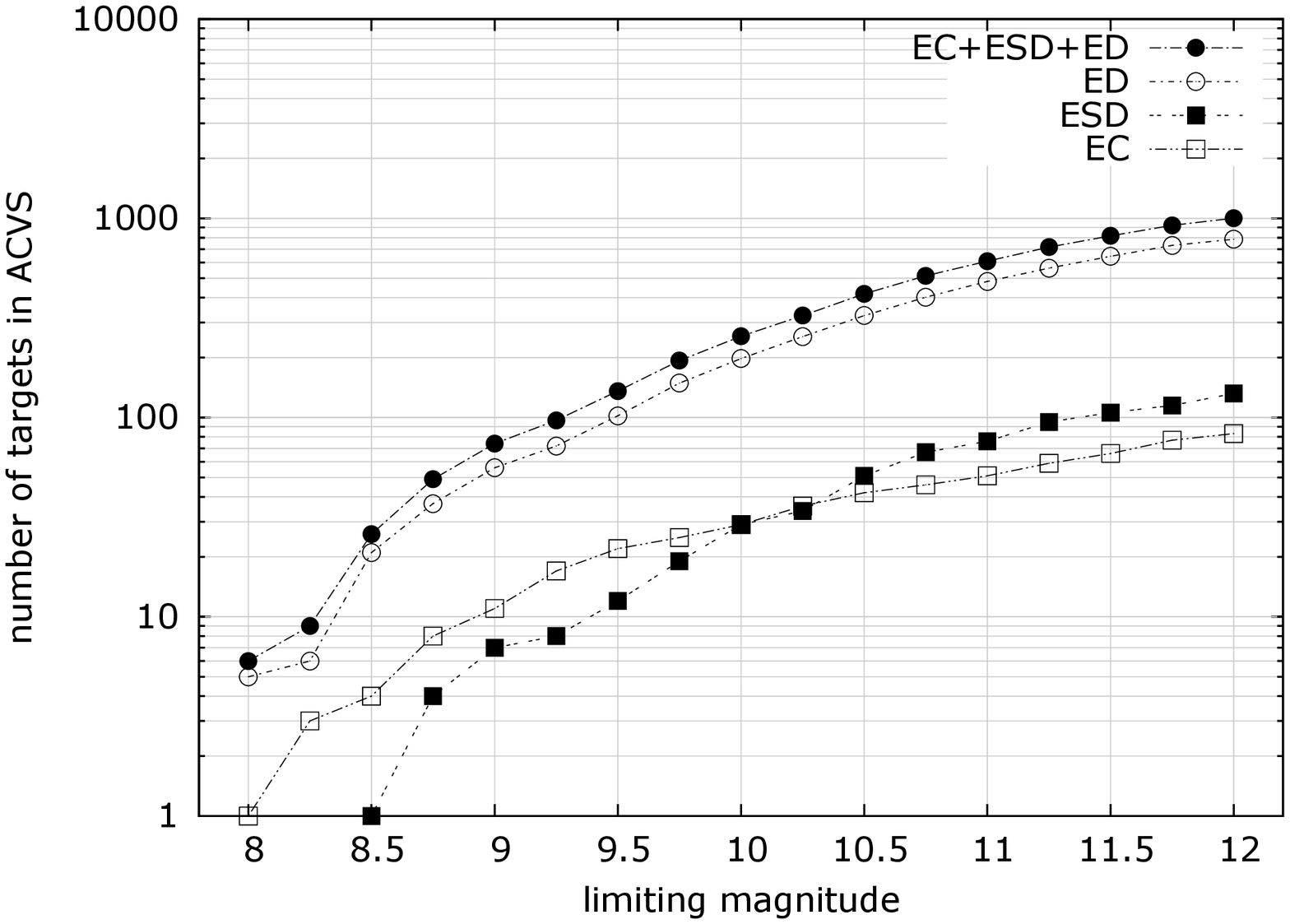}
\caption{Number of potential binary stars from the ACVS for given limiting magnitude and period longer than 3 days.}
\label{fig:ACVSstat}
\end{center}
\end{figure}

The ASAS Catalogue of Variable Stars (ACVS) \citep{Paczynski2006} includes 5,384 eclipsing contact binaries (EC), 2,957 semi-detached systems (ESD) and 2,758 detached systems (ED). ACVS is a very valuable source of potential targets for a spectroscopic survey. Limiting the brightness to 10 mag gives a total of 776 systems, comprising 231 EC, 143 ESD and 402 ED binaries (Fig. \ref{fig:ACVSstat}). Increasing the magnitude limit to 10.5 mag increases the sample to 1,291 binaries (395 EC, 229 ESD, and 667 ED). If we exclude fast orbiting systems to avoid binaries with wide spectral lines that may be potentially useless for radial velocity measurements and consider only targets that have periods of 3 days or more, we still get a total number of 418 targets. 
ACVS data has been been used to conduct spectroscopic surveys. Some of the recent publications include \cite{Parihar2009}, who conducted a spectroscopic survey of 180 binary systems aimed at the analysis of stellar activity. Spectral data was obtained with low resolution instruments ($R\sim1,600 - 3,600$).  \cite{Helminiak2009} present a survey of 18 binaries from the ACVS conducted using high resolution spectrographs ($R\sim40,000 - 60,000$), including spectra obtained with an iodine cell. The survey is ongoing and will cover a much wider set of systems. A large-scale, fully automatic spectroscopic survey of binaries from the ACVS has not yet been conducted, however.

\section{BAsic \'{e}CHElle Spectrograph -- BACHES}
\label{sec:BACHES}

BACHES\footnote{BAsic \'{e}CHElle Spectrograph, \textit{baches} means pothole in Spanish.} is a compact, lightweight, and inexpensive medium resolution ($R\sim20,000$) \'{e}chelle spectrograph manufactured by Baader Planetarium GmbH in accordance with a technology transfer license deal with the Max-Planck-Institut f\"{u}r Extraterrestrische Physik in Garching. The spectrograph has been designed by the Club of Amateurs in Optical Spectroscopy (CAOS) group that has conducted many successful instrumental projects in the past, e.g. DADOS, FIASCO, LECHES \citep{Mugrauer2009}. \cite{Avila2007} describe the internals of the spectrograph as well as results of initial tests of the instrument conducted in Paranal on a 0.35-m Celestron telescope, where it was mounted directly in the cassegrain focus of the telescope, without intermediate devices.

\subsection{The instrument}

The unit tested by us was a fully functional pre-production instrument having the same mechanical and optical characteristics as the final production version (at the time of the tests production of the first units was on the way). The spectrograph body is 290x100x52 mm in size and weighs less than 1.5~kg. BACHES' head extends by an additional 85 mm and houses the actual slit, flip-mirror mechanism with control cable connector, calibration fibre input port, a 2-inch cylinder that connects to the telescope and a slit-view/guider camera interface. All interfaces have standard dimensions and are mechanically compatible with amateur telescopes' components. BACHES' head allows one to acquire spectra of the actual object of interest as well as calibration spectra. In case of the latter, light is fed to the instrument via a calibration fiber. The remotely controlled flip-mirror is used to direct the proper light beam on to the slit.  We have tested the spectrograph working with the 25x100 $\mathrm{\mu m}$ slit. It is engraved in a reflective nickel plate and mounted in BACHES' head. Internally, the instrument consists of a doublet lens that collimates the light beam on to a 63 l/mm 73\degree \'{e}chelle grating. Then a diffraction grating cross-disperses the beam that is projected on the CCD chip using an objective. The instrument is optimised for f/10 input beams and cameras with 1530x1020 $9~\mu m$ pixels \citep{Avila2007}. Further, detailed information about the internals of instrument is unavailable due to the commercial nature of this product.

\subsection{Baader Spectroscopy Remote Calibration Unit}

The spectrograph is provided with a calibration device called Baader Spectroscopy Remote Calibration Unit (RCU). The RCU consists of a power supply, a Thorium-Argon lamp, a halogen lamp and a web-enabled remote control interface. The remote control interface allows one to remotely switch the lamps on and off, as well as control the flip mirror's position in BACHES' head. Light from the calibration lamps is directed to a 2.5-m 50 $\mathrm{\mu m}$ fibre that is connected the spectrograph. The RCU is controlled via a web browser interface that allows easy integration with other components of the system. Additionally, the RCU allows the user to manually control the flip-mirror and the calibration lamps using a control panel on the device.

\section{Test site description}
\label{sec:TestSite} 

\begin{figure*}
\begin{center}
\includegraphics[width=\textwidth]{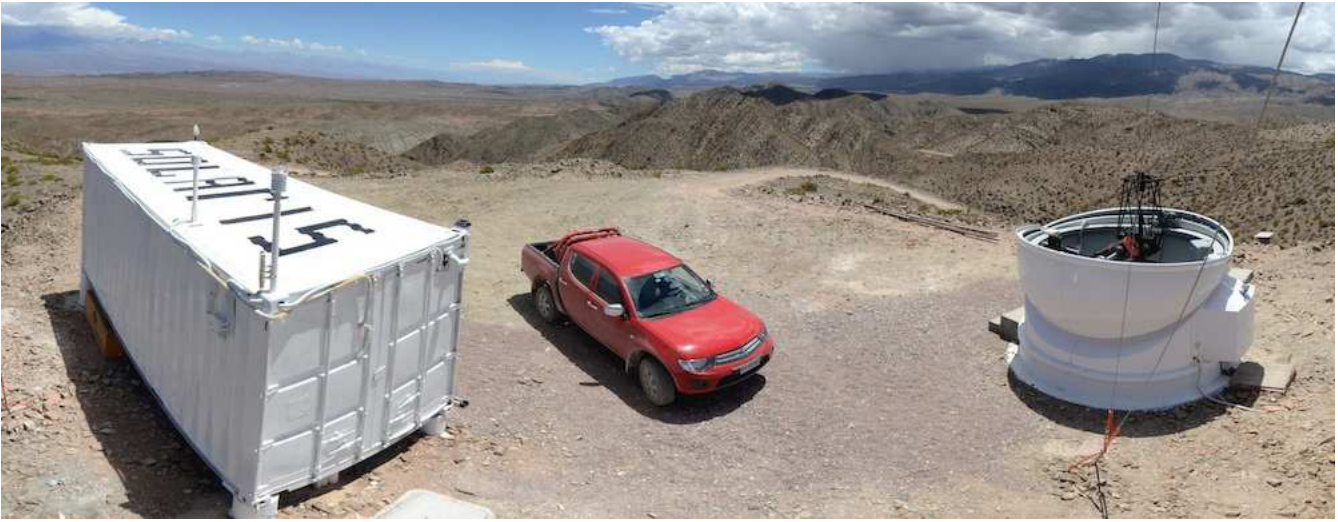}
\caption{Solaris-4 observatory in CASLEO, Argentina during the commissioning phase.}
\label{fig:CASLEOPano}
\end{center}
\end{figure*}

Project Solaris is a Polish scientific initiative to open a new frontier in extrasolar planets hunt \citep{Kozlowski2013}. It consists of four autonomous observatories located in the Southern Hemisphere: Solaris-1 and Solaris-2 in the South African Astronomical Observatory in South Africa,  Solaris-3  in Siding Spring Observatory in Australia and Solaris-4 in Complejo Astron\'omico El Leoncito (CASLEO) in Argentina\footnote{http://www.casleo.gov.ar}. CASLEO is located in the San Juan Province, near the town Barreal. It is the biggest Argentinian observatory, that among others, hosts a 2.15~m telescope. The Solaris-4 site is 7 km (by road) away from the main observatory buildings, on an isolated mountain top (S 31$\degree$47'3''  W 69$\degree$18'17'', elev. 2635 m AMSL). The telescope is housed in a Baader Planetarium 3.5-m AllSky Dome fitted with a custom building management system and additional security systems. A 20-ft office container located 10 m from the dome stores the main rack with the control computer, networking devices, media converters and other auxiliary components (Fig. \ref{fig:CASLEOPano}). A smaller rack is located in the dome and is equipped with the necessary power supplies and networking devices. Data and control signals to/from the instruments are transferred via three fibre links that connect the dome with the container. Both the dome and the container are equipped with independent Uninterruptible Power Supply (UPS) systems. The observatory is fitted with a range of sensors and receivers that allow it to be operated remotely or autonomously: two weather stations, lightning detector, GPS receiver, surveillance cameras, all-sky camera. Tests were conducted during the final days of the commissioning phase of the entire Solaris-4 observatory at the turn of November and December 2013. The Solaris Network is dedicated to colour photometry and the detection of circumbinary planets using the eclipse timing technique. The Solaris-4 telescope has been used as the test bed for the spectrograph during its evaluation phase.

\subsection{Telescope}

Solaris-4 is a 0.5-m diameter f/15 Ritchey--Chr\`etien optical system housed  in a carbon-fibre truss design optical tube assembly (OTA). The telescope is equipped with a motorised focuser and motorised primary mirror covers. The OTA is mounted on a Astrosysteme Austria (ASA) DDM160 modified German Equatorial mount equipped with direct drive motors and high resolution incremental encoders.

\subsection{Imaging train components}

\begin{figure}
\begin{center}
\includegraphics[width=\columnwidth]{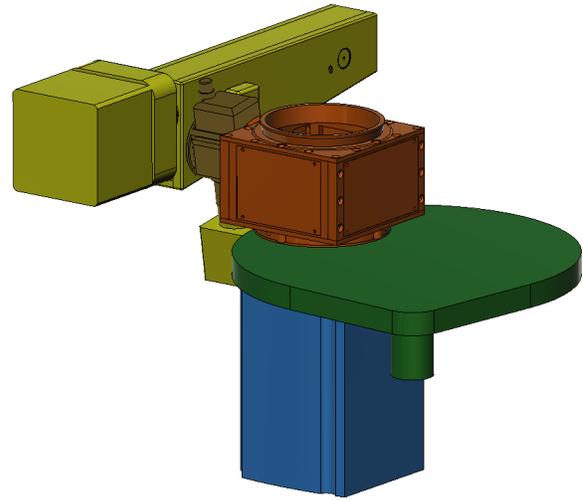}
\caption{Engineering model of the Solaris-4 imaging train: BACHES with the main camera and slit-view camera (yellow), GAM (red), filter wheel (green) and photometric CCD camera (blue).}
\label{fig:GAMPorts}
\end{center}
\end{figure}

\begin{figure}
\begin{center} 
\includegraphics[width=\columnwidth]{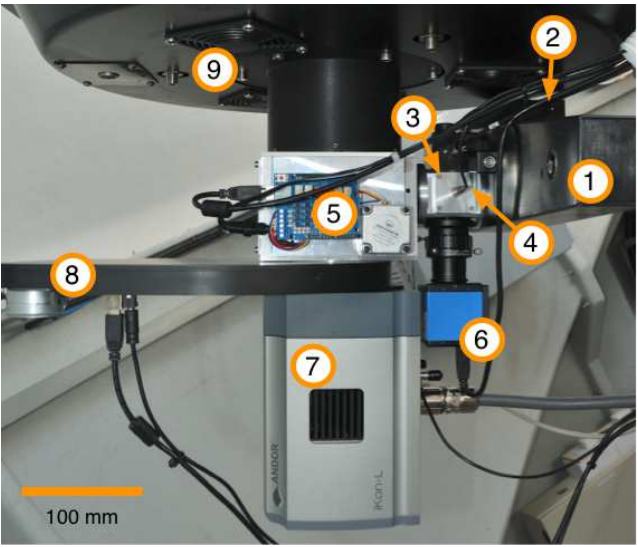}
\caption{Imaging train: BACHES spectrograph (1), SBIG CCD camera (2), BACHES head (3), calibration fibre (4), GAM (5), slit view/guide camera (6), photometric CCD camera (7), filter wheel (8), telescope back plate (9).}
\label{fig:ImagingTrain}
\end{center}
\end{figure}

\begin{figure}
\begin{center}
\includegraphics[width=\columnwidth]{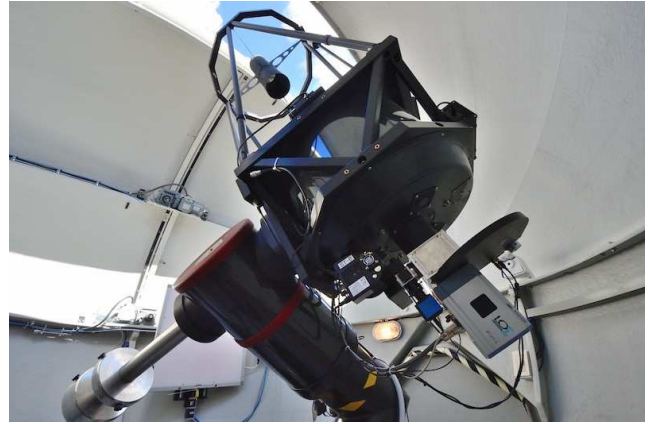}
\caption{Solaris-4 test-bed setup: ASA mount, OTA and imaging train with photometric camera and spectrograph.}
\label{fig:ImagingTrainTelescope}
\end{center}
\end{figure}

The original imaging trains installed on all Solaris telescopes include: a field rotator (manufactured by ASA), a 12-position filter wheel (Fingerlakes Instruments CFW-12) equipped with 50 mm round Johnson and Sloan filter sets and a CCD camera (Andor iKon-L 936). For BACHES' tests, the imaging train has been modified to accommodate the additional equipment. Due to the mechanical constraints of the system, the field rotator has been removed and replaced with a guide and acquisition module (GAM) which has been designed in such a way that the back focus distance to the CCD camera remained unchanged and the position of the secondary mirror (focus) for both photometry and spectroscopy was identical. An engineering model of the imaging train is shown in Fig. \ref{fig:GAMPorts} and the actual image of the setup is shown in Figs. \ref{fig:ImagingTrain} and \ref{fig:ImagingTrainTelescope}.

\subsection{Guiding and Acquisition Module}

\begin{figure}
\begin{center}
\includegraphics[width=\columnwidth]{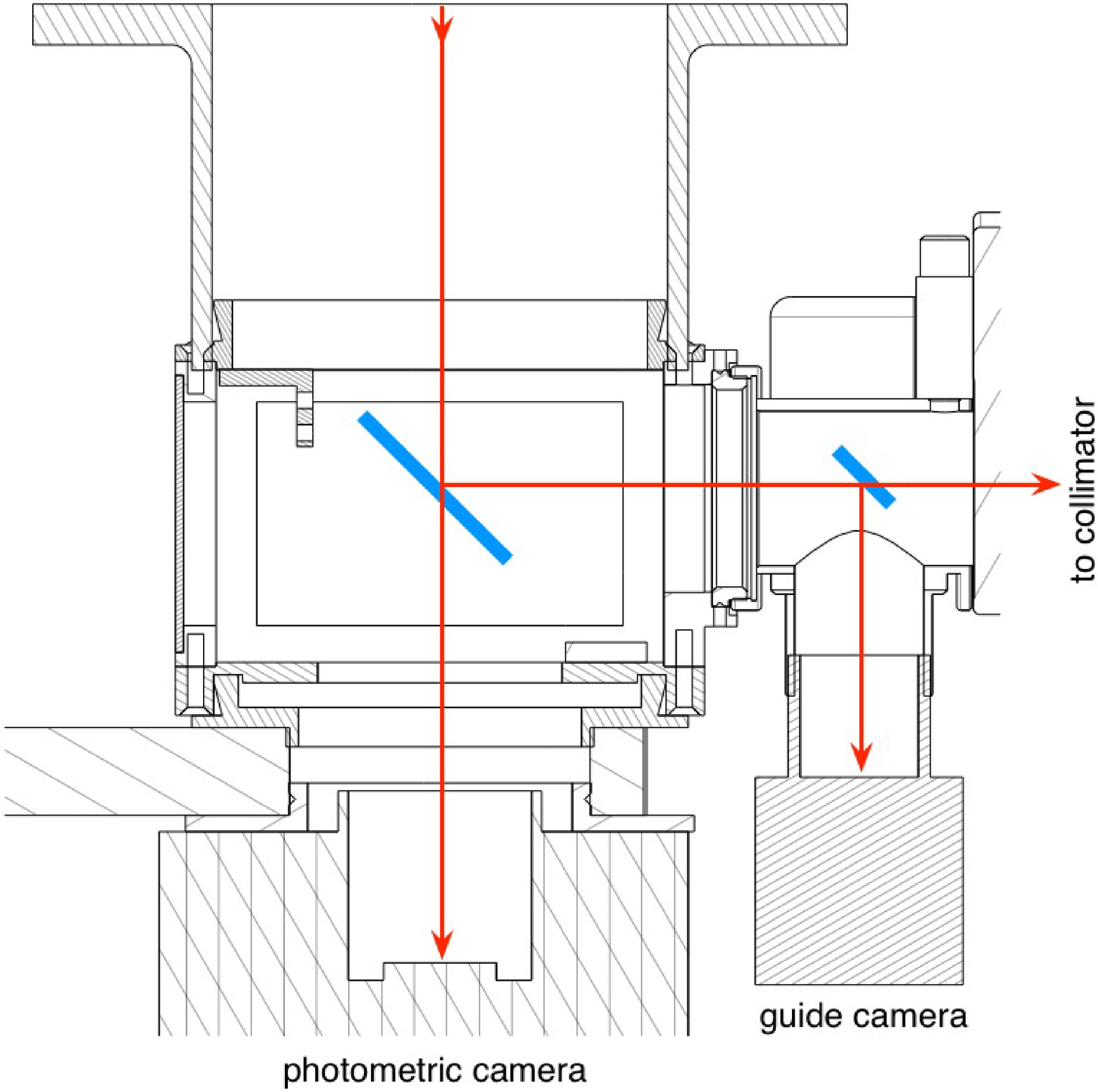}
\caption{GAM ports configuration. Blue rectangles indicate mirrors.}
\label{fig:LightPath}
\end{center}
\end{figure}

A GAM has been designed specifically for BACHES' tests. It is a modular construction that has one optical input port and two optical output ports (Fig. \ref{fig:LightPath}). The input port has a 4" ASA dovetail interface that allows it to be attached to the telescope with locking screws. The straight-through port has a 3 inch 24UNS-2B male thread that screws into the corresponding female thread of the filter wheel. The third port is a standard 2 inch eyepiece port with three side locking screws. The depth of the port has been optimised for BACHES. Switching between the output ports is accomplished with a motorised flip-mirror. The mirror itself is a Thorlabs BBSQ2-E02 2x2 inch Square Broadband Dielectric Mirror with reflectivity above 99\% in the entire 400-750 nm wavelength range. The flip-mirror mechanism's rotary joint consists of an aluminium mount with two precision ball bearings, a steel shaft and the mirror support plate. The shaft is driven by a stepper motor via an aluminium-nylon coupling. The mirror is glued to the support using a 0.25 mm thick double sided acrylic tape. To assure high precision positioning of the mirror keeping the design simple and cheap to manufacture, a spring mechanism has been implemented. It allows the windings of the stepper motor to be not energised when the mirror is stationary thus reducing the heat generated by the motor to a negligible level even in case of frequent position switching. Limit screws are used to control the end positions of the mirror support plate. All components, unless otherwise noted, are made of aluminium and have been CNC-machined based on a professional three-dimensional engineering model. The GAM has a box-like structure consisting of six plates and two covers. The plates are screwed together using 44 M4 bolts to form a rigid structure. The 1.8$\degree$/step stepper motor is driven by an AVR-based\footnote{AVR is a series of micro-controllers manufactured by Atmel Corp.} controller. Communication with the PC is established via a USB link and is realised using a custom application that allows to control the position of the flip-mirror.

\subsection{CCD and slit-view/guide camera}

An SBIG ST-8XMEI camera was used as the main imaging camera for testing purposes. It has a peak QE of nearly 85\%, and incorporates the KAF-1603ME chip with 1530 x 1020, 9~$\mathrm{\mu}$m square pixels. An Imaging Source DMK 21AU04.AS with 640x480 5.6 micron square pixels was used as the slit-view and guide camera. The ST-8 camera was attached to BACHES using a focusing adapter and a 2 to 1.25-inch reducer. This temporary setup, though good enough for tests, exhibited significant flexure visible between successive spectra. The same connection interface was used for the DMK camera and performed much better mainly due to the low weight of the camera.

\subsection{Power supply and data/control cables}
Components of the imaging train are usually connected to the ports available on the declination axis of the mount so that they do not restrict the movements of the telescope. This is an important feature of remote/autonomous operation of the telescope. In case of BACHES, however, the cables have been led externally, on the mount's enclosure. This was necessary due to the number of additional signal and power cables that are required to use the spectrograph. The fibre from the RCU has been additionally secured by a flexible cable trunking. A USB hub has been used to multiplex signals from the cameras and the GAM that could then be transferred to the main control computer using a fibre connection.

\section{Software}

The software component of an autonomous observatory is a key element that allows the hardware to be used in the most effective and safest manner. During the tests of the spectrograph the observatory was operating in a remote mode (instead of an autonomous mode) due to the addition of new hardware components that have not been incorporated into the system yet in the form of appropriate software drivers and modules. During all but the first two nights, observations were carried out from the main observatory buildings (7 km away) via a wireless link to the Solaris-4 site.

\subsection{Control and acquisition software}

\begin{figure}
\centering
\includegraphics[width=\columnwidth]{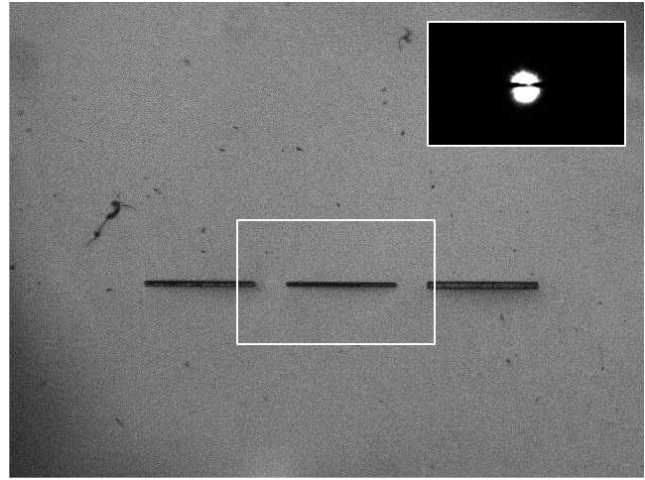}
\caption{DMK guide camera image showing three slits -- from left: 75, 25 and 100 $\mathrm{um}$. Inset is a to-scale image of the star centred on the 25 $\mathrm{um}$ slit. Due to low, 8-bit depth of the image and the high contrast, the star appears overexposed.}
\label{fig:DMK}
\end{figure}

\textsc{maximdl}\footnote{http://www.cyanogen.com} was used as the primary camera control, image acquisition and guiding software. It performed very well with the ST-8 camera. The \textsc{maximdl} driver for the DMK slit-view camera is not very capable. It provides very basic functionality that is accessible to the user in a rather inconvenient way that is especially troublesome when observing remotely over a low-bandwidth connection. The driver forces the image from the slit-view camera to be streamed rather than taking single exposures that would be more adequate in this application. Nevertheless, the sensitivity of the guide camera is good enough to guide on 10 mag stars with exposure times below 1 sec. Guiding itself performed well, though sometimes \textsc{maximdl} struggled to properly position the star on the slit. The axis along the width of the slit is more problematic because the slit cuts the stellar image in half. Eventually, a dedicated software guide module needs to be developed that will work correctly on non-gaussian stellar profiles. Sample images obtained with the slit-view camera are showed in Fig. \ref{fig:DMK}.

\subsection{Image reduction and analysis pipeline}
The spectra were reduced with the standard \textsc{echelle} package from \textsc{iraf}\footnote{\textsc{iraf} is written and supported by the \textsc{iraf} programming group at the National Optical Astronomy Observatories (NOAO) in Tucson, AZ. NOAO is operated by the Association of Universities for Research in Astronomy (AURA), Inc. under cooperative agreement with the National Science Foundation. http://iraf.noao.edu/}. The wavelength calibration was based on a pair of Th-Ar exposures one taken before and the other after a target's exposure. The radial velocities (RVs) were computed using the RVSAO package by performing a cross correlation of the observed spectra with synthetic stellar spectra. For binary stars the RVs were computed using the TwO Dimensional CORrelation technique (TODCOR) \citep{Zucker1994}. TODCOR allows one to compute RVs of double-lined spectroscopic binaries based on a pair of cross correlation functions computed with two synthetic template spectra matching the components of the binary.     

\section{Test results}
\label{sec:TestResults} 
In the following sections we present the results of mechanical and scientific tests conducted with the hardware and software setup describe above.
\subsection{Mechanical tests}
The setup described in this paper, as noted earlier, is a dual facility configuration that is designed for remote operation. Both modes of operation, i.e. photometry and spectroscopy, are available to the user at any time. We have conducted two types of mechanical stability tests. The first one tests the stiffness of the imaging train to determine the flexure of the instruments relative to the optical axis of the telescope. Effectively, this test shows two things: (1) how good is the design of the GAM unit in terms of its mechanical stiffness and (2) how much does the flexure influence pointing precision - this will be crucial in future automatic operation where the object of interest will have to be centred on the slit. Mechanical tests of the imaging train's stiffness were performed in the following routine. First the telescope was pointed at the zenith, the star was centred on the photometry camera and the position of the star in the slit-view camera was recorded. This cycle has been repeated for N, E, S and W horizon points (25 degrees above the horizon). The positions of the star on the images obtained with the slit-view camera have been measured. The results are presented in Table \ref{tab:FlexureTable}. The values have been corrected according to the meridian flip of the mount. It is clearly visible that imaging train does exhibit flexure. This does not directly influence spectroscopy, but should be considered as something to be modified in the next version of the GAM to optimise pointing precision in the automatic mode. The most significant input to the flexure errors is introduced by the way the Andor CCD camera and the filter wheel assembly are attached to the GAM. A revised and updated version of this interface will limit flexure at this point and the reduce the overall flexure of the imaging train. 
\begin{table*}
\begin{center}
\begin{tabular}{ccccccccccc}
\hline
\specialcell{telescope\\ pier side}	& position	& correction	& $x$ (px) &	$y$ (px) &$\Delta x$ (px) &	$\Delta y$(px)	&	\specialcell{$\Delta x$\\ ($\mu$m)}	&	\specialcell{$\Delta y$ \\($\mu$m)}	&	\specialcell{total flexure \\($\mu$m)} &	\specialcell{total flexure \\(arc sec)}\\ \hline \hline
E	&	Zenith	&	1	&	313	&	337	&	0	&	0	&	0.0	&	0.0	&	0.0	&  0.0 \\ 
E	&	N	&	1	&	321	&	327	&	-8	&	10	&	-44.8	&	56.0	&	71.7 & 9.9	\\
W	&	E	&	-1	&	311	&	349	&	-2	&	12	&	-11.2	&	67.2	&	68.1	& 9.4\\ 
W	&	S	&	-1	&	318	&	354	&	5	&	17	&	28.0	&	95.2	&	99.2	 & 13.6\\ 
E	&	W	&	1	&	291	&	350	&	22	&	-13	&	123.2	&	-72.8	&	143.1 & 19.7	\\ \hline
\end{tabular}
\end{center}
\caption{Flexure test results. $x$ and $y$ are pixel positions of the star on the guide camera, $\Delta x$ and $\Delta y$ are the differences in the positions in the respective axes compared to the position at the zenith. In all five cases the star's image was centered on the photometry CCD camera.}
\label{tab:FlexureTable}
\end{table*}

\begin{figure}
\begin{center}
\includegraphics[bb=  44 0 734 513, clip, width=\columnwidth]{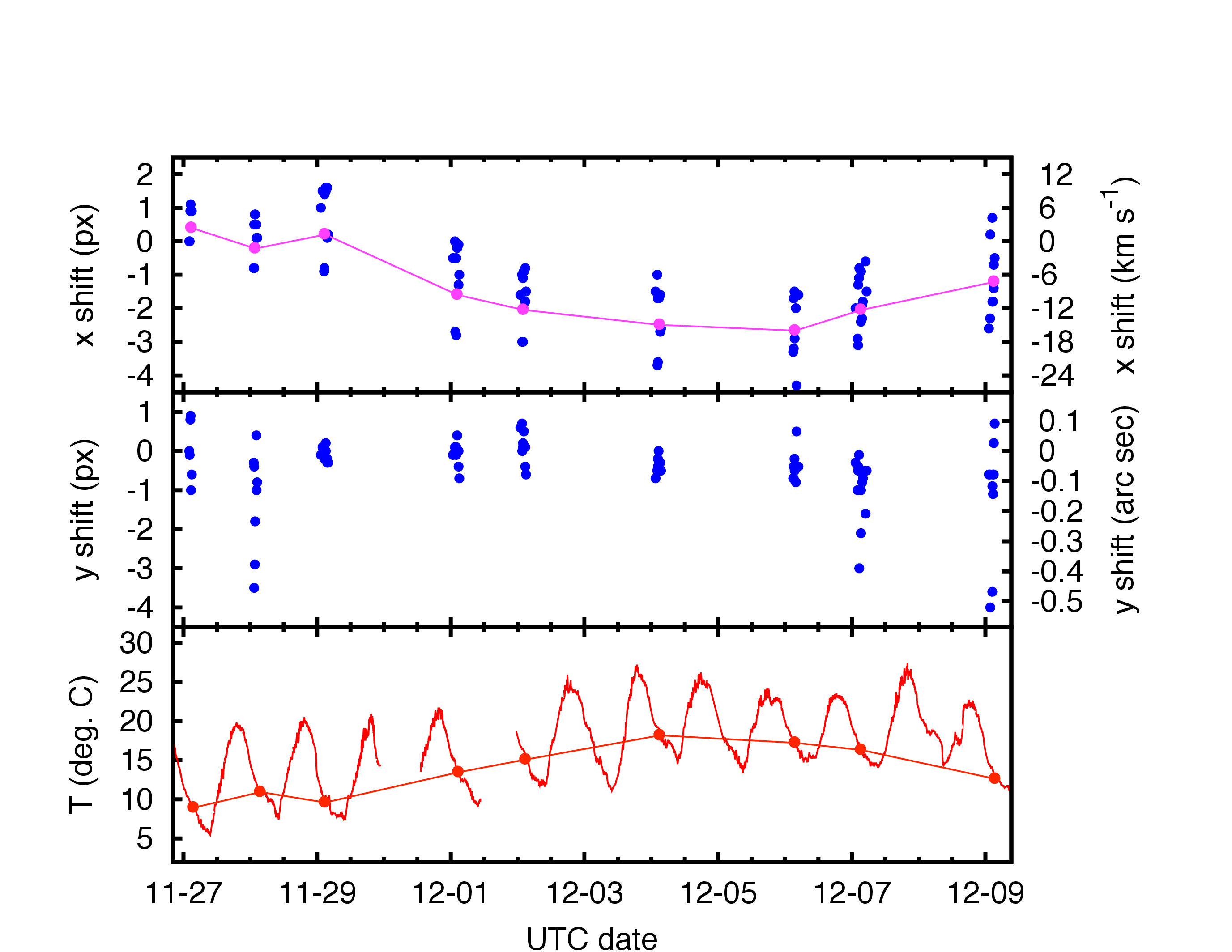}
\includegraphics[bb= 213 58 448 735, angle=-90, clip, width=\columnwidth]{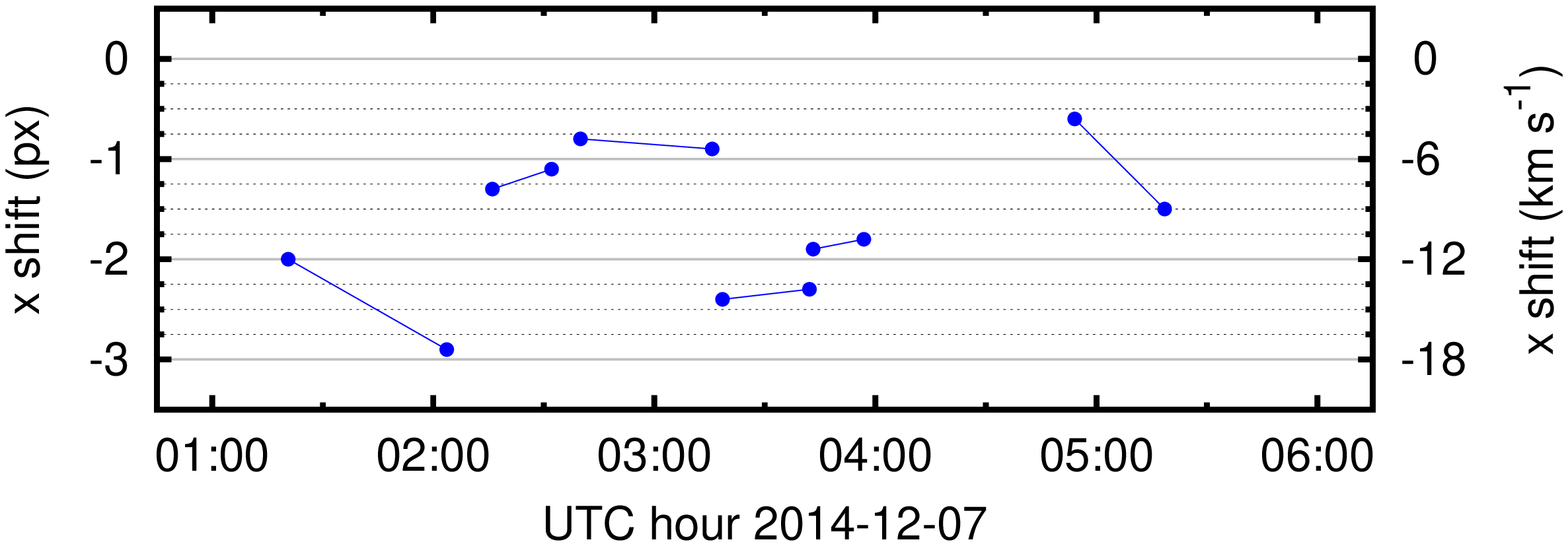}
\caption{Top panel: shifts between all acquired Th-Ar frames and the first Th-Ar frame from Nov 26th in $x$ and $y$ directions in pixels, km~s$^{-1}$ and arc secs. $x$ is the dispersion direction. 1 pixel corresponds to $\sim6$ km~s$^{-1}$ in radial velocities. Shifts in the dispersion direction have been averaged for each night (magenta points) and projected onto the ambient temperature at corresponding times (red points). Bottom panel: shifts during the observing night of Dec 7th. Point pares indicate Th-Ar exposures taken before and after object spectra. It should be noted that light from the Th-Ar calibration lamp is injected directly into the spectrograph's head, thus it is independent of the optics of the telescope and GAM. However, since the length of the slit (100 $\mu$m or 2.77 arc sec on sky) projects to the height of the \'{e}chelle order, the shifts in $y$ can be expressed in arc secs.
}
\label{fig:flexure}
\end{center}
\end{figure}

The second type of mechanical tests shows the actual stability of the spectrograph. In this case we have measured the shifts between the spectra of the Th-Ar  lamp acquired throughout the observing runs relative the the first Th-Ar spectrum captured during the first observing night. The obtained results are sensitive to the mechanical coupling of the spectroscopy CCD camera with the spectrograph as well as to the intrinsic properties of the instrument itself that is subject to changing ambient conditions such as temperature and humidity and internal and external disturbances caused by mechanical stress, e.g. wind and changing orientation of the instrument relative to the gravity vector. Changing ambient conditions affect not only the mechanics of the instrument but also the refractive index of air that is enclosed in the instrument. The results are shown in Fig. \ref{fig:flexure}. Though shifts as big as 5 pixels (30 km~s$^{-1}$) throughout the entire test period can be observed, the pairs of Th-Ar calibration spectra taken before and after the object spectra show much better stability. Shifts between the connected pairs are mainly caused by the flexure of the spectrograph-camera coupling that becomes apparent when the telescope points at different positions on the sky. We have checked for correlations between various parameters describing local ambient conditions and we have found that the long term shifts of the spectra in the dispersion direction are correlated with the changing ambient temperature which is clearly visible on the plot. Short term stability of the instrument is better than a fractions of a pixel rms, what is confirmed by scientific results presented in \ref{ssec:Binaries}. Thanks to the Th-Ar calibration lamp, most of the long and short term instabilities of the spectrograph can be removed.

The flip-mirror system behaved properly and the "spectroscopy" position of the mirror showed good repeatability. After successive changes of the mirror's position (5 minutes total, flip every 10 seconds), the star's image would land on the same pixels of the slit-view camera (as permitted by seeing and tracking errors).

\subsection{Observing runs}

\begin{figure}
\begin{center}
\includegraphics[width=\columnwidth]{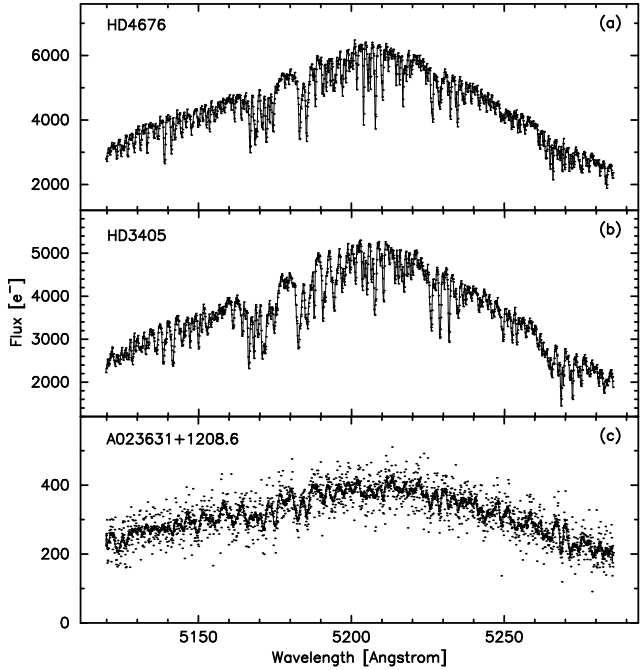}
\caption{An \'{e}chelle order from BACHES for three binary systems HD4676, HD3405 and A023631+1208.6.
The epoch of the spectra is such that the radial velocity differences between the components of the
binaries are large and one can see the magnesium triplet from both components. The dots in the panel (c) represent the actual recored flux of A023631+1208.6. The solid line represents a smoothed spectrum (boxcar smoothing with 10 pixels).}
\label{fig:Spc}
\end{center}
\end{figure}

A total of 7 objects have been observed from Nov 26th to Dec 8th 2013: spectroscopic standard stars and binary stars. The goal was to check the stability of the spectrograph using standard stars and to verify the ability to characterise binary systems using radial velocities following a scheme that will very likely be adopted in a future spectroscopic survey. Observed objects have been listed in Table \ref{tab:Objects}. An \'{e}chelle order containing the magnesium triplet is shown in Fig.~\ref{fig:Spc} for three binary stars HD4676, HD3405 and A023631+1208.6. On average the seeing was 1.2 arc sec, stable. All observations were carried out when wind speed was below 30~km~h$^\text{-1}$, 15~km~h$^\text{-1}$ on average.
\begin{table*}
\begin{center}
\begin{tabular}{cccccccc}
\hline
 date & \small HD10700 &  \small HD1581 &  \small HD4676 &  \small HD3405 &  \small A023631+1208.6 &  \small HD75289 &  \small A034413-4116.8 \\ \hline \hline
26 Nov & 1 & 1 & 1 & - & - & - & - \\ 
27 Nov & 1 & 1 & 1 & 1 & - & - & - \\ 
28 Nov & 1 & 1 & 1 & 1 & 1 & - & - \\
29 Nov & - & - & - & - & - & - & - \\
30 Nov & 1 & 1 & 1 & 1 & 1 & - & - \\ 
1 Dec & 1 & 1 & 1 & 1 & 1 & - & - \\ 
2 Dec & - & - & - & - & - & - & - \\ 
3 Dec & 1 & 1 & 1 & 1 & 1 & - & - \\
4 Dec & - & - & - & - & - & - & - \\ 
5 Dec & 1 & 1 & 1 & 1 & - & 1 & - \\ 
6 Dec & 1 & 1 & 1 & 1 & 1 & - & 2 \\ 
7 Dec & - & - & - & - & - & - & - \\ 
8 Dec & 1 & 1 & 1 & - & 1 & - & 1 \\ 
remark & single & single & binary & binary & binary & planet & binary \\
mag & 3,5 & 4,2 & 5,1 & 6,8 & 9,8 & 6,4 & 8,94 \\ 
exp. (s) & 300 & 600 & 600 & 1200 & 1800 & 1800 & 1200 \\ \hline

\end{tabular}
\end{center}
\caption{Objects observed with BACHES and Solaris-4. Number indicates number of spectra acquired per object.  Days with no observations mean bad weather (mostly) or technical problems (unrelated to the spectrograph, i.e. network was down). }
\label{tab:Objects}
\end{table*}%

\subsection{Calibration}
\begin{figure}
\subfloat[]{\includegraphics[width= \columnwidth]{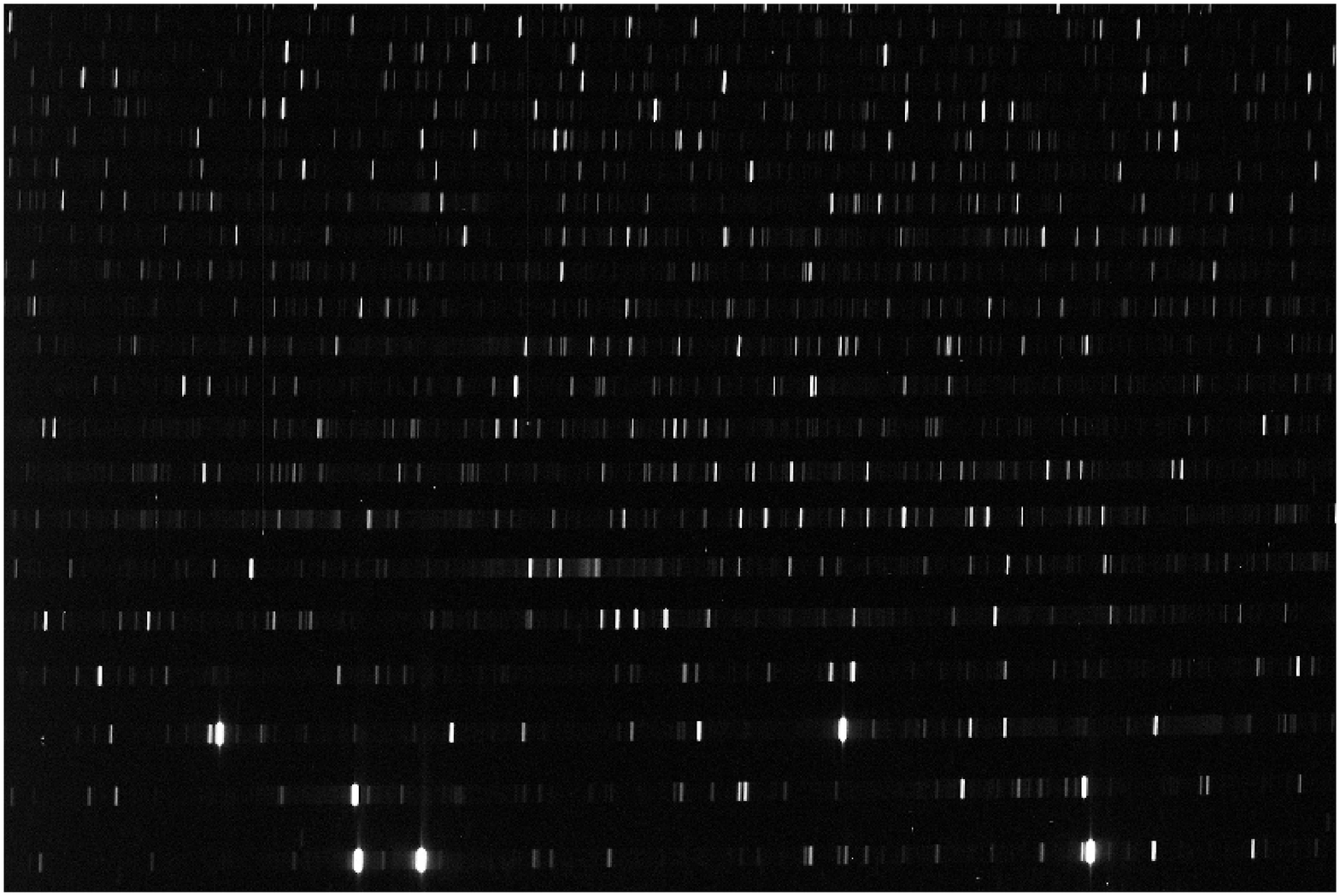}\label{fig:Th-Ar}}\\
\subfloat[]{\includegraphics[width= \columnwidth]{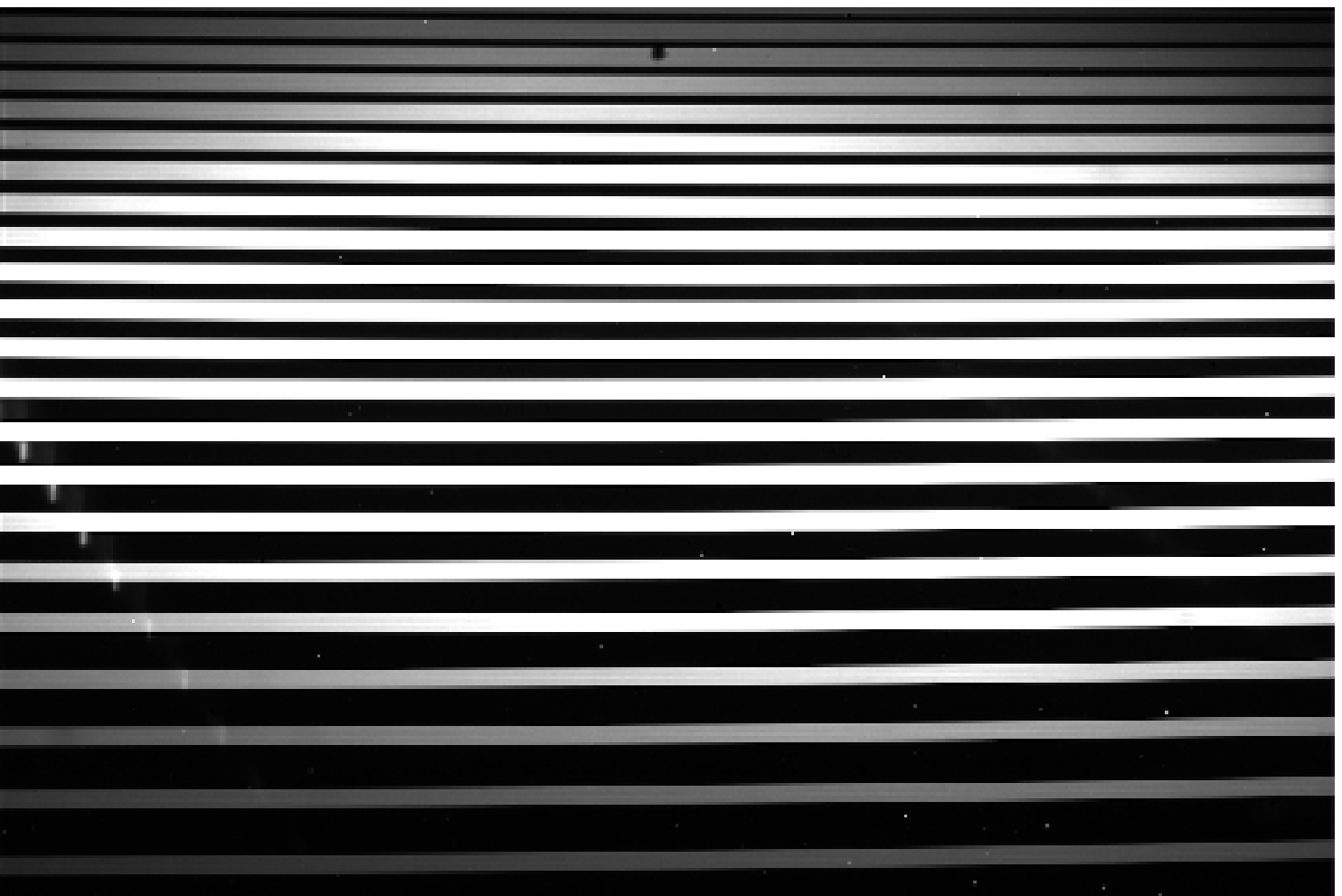}\label{fig:Flat}}\\
\subfloat[]{\includegraphics[width= \columnwidth]{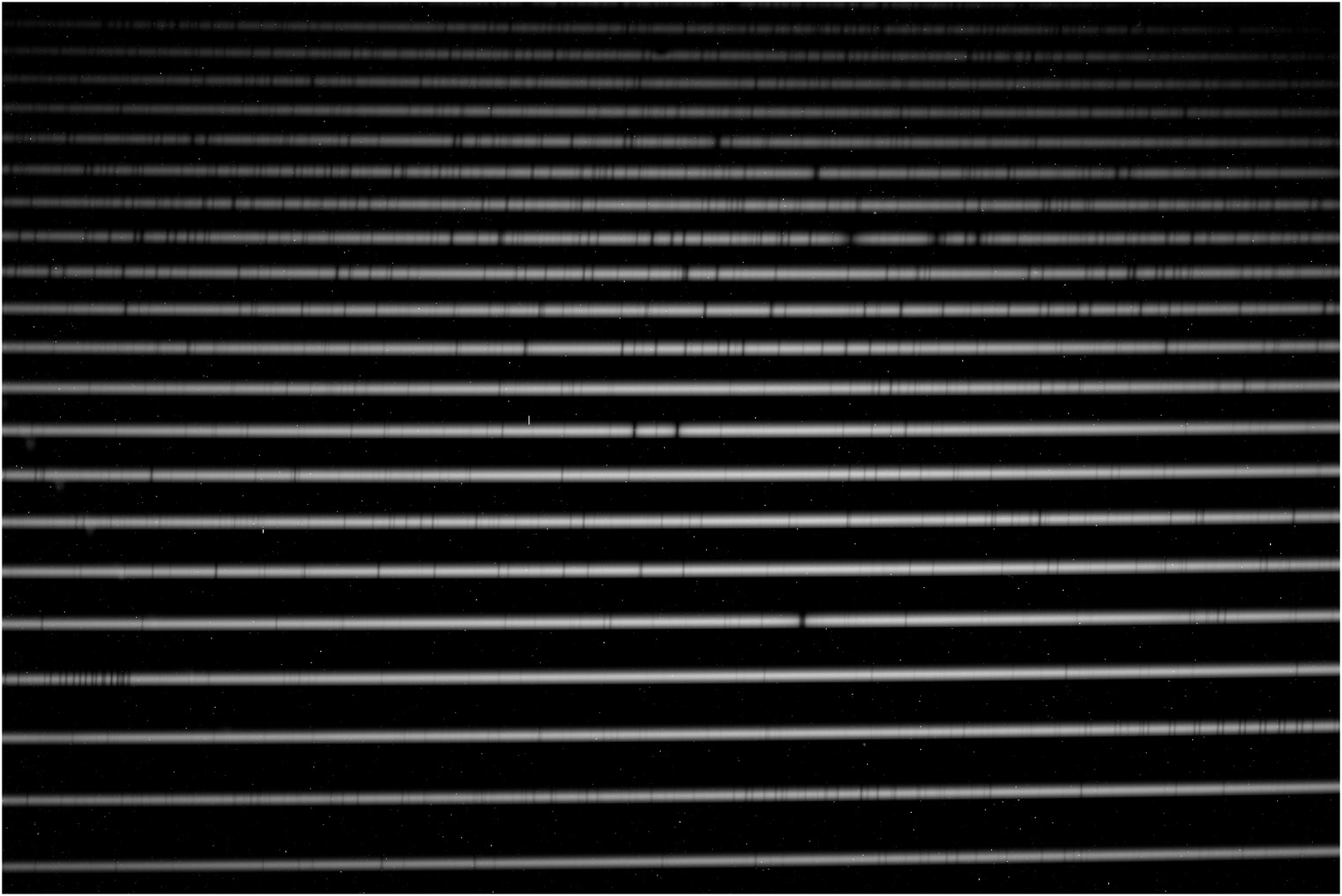}\label{fig:HD10700frame}}\\
\caption{Frames showing 1 sec Th-Ar (a), 60 s flat lamp (b) and  300 s HD10700 (c) raw spectra.}
\end{figure}

A series of flats has been taken every night, dark and bias frames only on Nov 30th. Th-Ar spectra have been taken immediately before and after the target exposure. Th-Ar lines have been manually identified using a spectroscopic atlas and IRAF. Flexure of the SBIG camera relative to the spectrograph introduced shifts between successive Th-Ar spectra thus increasing the formal errors of the calculated radial velocities (RMS of ~1.5 km~s$^{-1}$). Exemplary Th-Ar and flat calibration spectra are shown in Figs. \ref{fig:Th-Ar} and \ref{fig:Flat}, respectively. With the described setup the obtained raw spectrum consists of 21 orders covering 440 to 758 nm. The width the orders varies from 14.1 to 23.6 nm. The 25 x 100 $\mu$m slit projects to 0.69 x 2.77 arc seconds on the sky.

\subsection{Standard Stars}

\begin{figure}
\begin{center}
\includegraphics[width=\columnwidth]{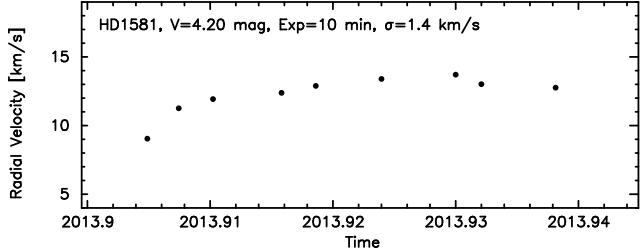}
\caption{Radial velocities of HD1581 as a function of time.}
\label{fig:HD1581-rv}
\end{center}
\end{figure}

\begin{figure}
\begin{center}
\includegraphics[width=\columnwidth]{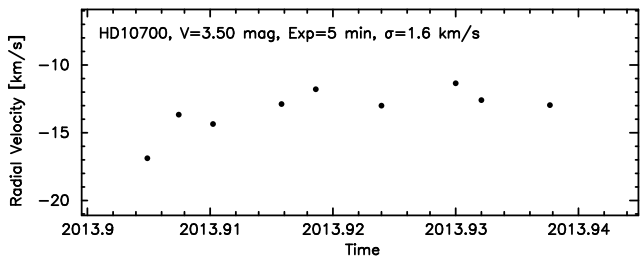}
\caption{Radial velocities of HD10700 as a function of time.}
\label{fig:HD10700-rv}
\end{center}
\end{figure}

To test the stability of BACHES, standard stars have been observed (HD1581, V=4.2 mag and HD10700, V = 3.5 mag). Resulting radial velocities are presented in Figs. \ref{fig:HD1581-rv} and \ref{fig:HD10700-rv}, an example frame with the spectrum of HD10700 is shown in Fig. \ref{fig:HD10700frame}. The results are consistent, though a trend of unclear origin is visible when comparing the two plots, despite the careful calibration with the Th-Ar lamp spectra that removes effects caused by the mechanical instability of the spectrograph.

\subsection{Binaries -- radial velocities}
\label{ssec:Binaries}

\begin{figure}
\begin{center}
\includegraphics[bb= 50 239 561 700,width=\columnwidth]{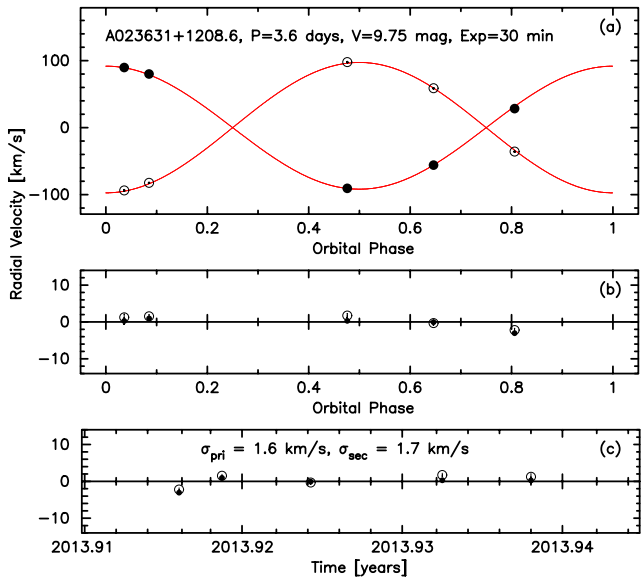}
\caption{(a) Radial velocities of an eclipsing binary star A023631+1208.6
as a function of the orbital phase. The best fit model is depicted with a solid line.
(b) Best-fit residuals as a function of the orbital phase. (c) Best-fit residuals as a function
of time.}
\label{fig:A023631+1208-rv}
\end{center}
\end{figure}

\begin{figure}
\begin{center}
\includegraphics[bb = 50 239 561 700, width=\columnwidth]{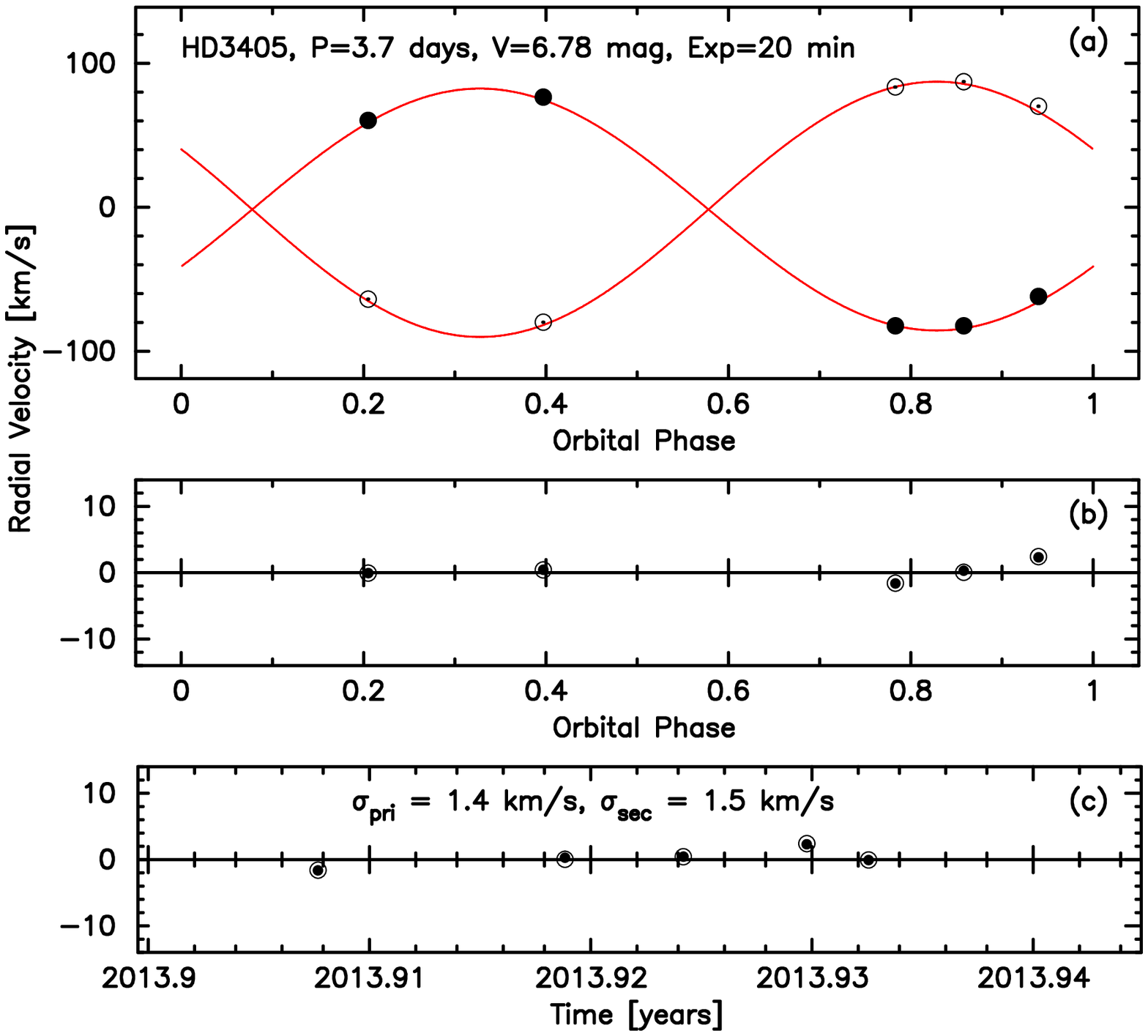}
\caption{(a) Radial velocities of a double-lined spectroscopic binary star HD3405
as a function of the orbital phase. The best fit model is depicted with a solid line.
(b) Best-fit residuals as a function of the orbital phase. (c) Best-fit residuals as a function
of time.}
\label{fig:HD3405-rv}
\end{center}
\end{figure}

\begin{figure}
\begin{center}
\includegraphics[bb = 50 239 561 700, width=\columnwidth]{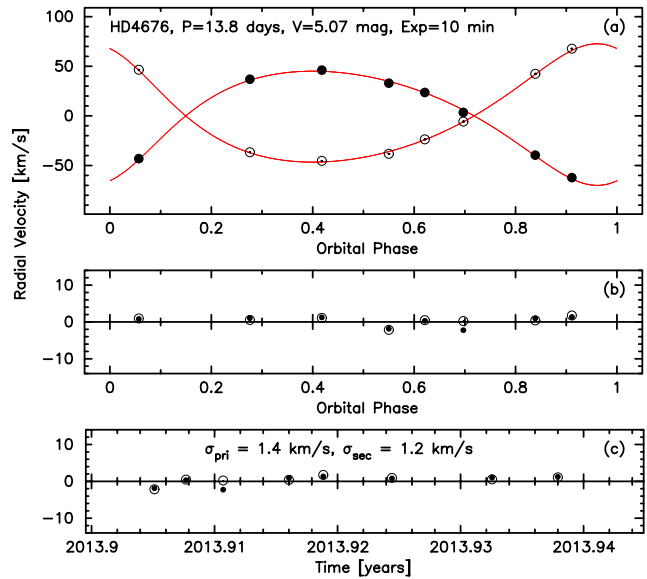}
\caption{(a) Radial velocities of a double-lined spectroscopic binary star HD4676
as a function of the orbital phase. The best fit model is depicted with a solid line.
(b) Best-fit residuals as a function of the orbital phase. (c) Best-fit residuals as a function
of time.}
\label{fig:HD4676-rv}
\end{center}
\end{figure}

The results of RV measurements have been shown in figs. \ref{fig:A023631+1208-rv} - \ref{fig:HD4676-rv}. For HD3405 and HD4676 we used the existing high precision orbital parameters and obtained the best-fit residuals by fitting only for an RV offset. For A023631+1208.6 we used the orbital period and T0 moment from the ASAS photometry and fitted for the RV amplitudes and the RV offset.

\begin{figure}
\begin{center}
\includegraphics[angle = -90, bb= 108 67 548 752, width=\columnwidth]{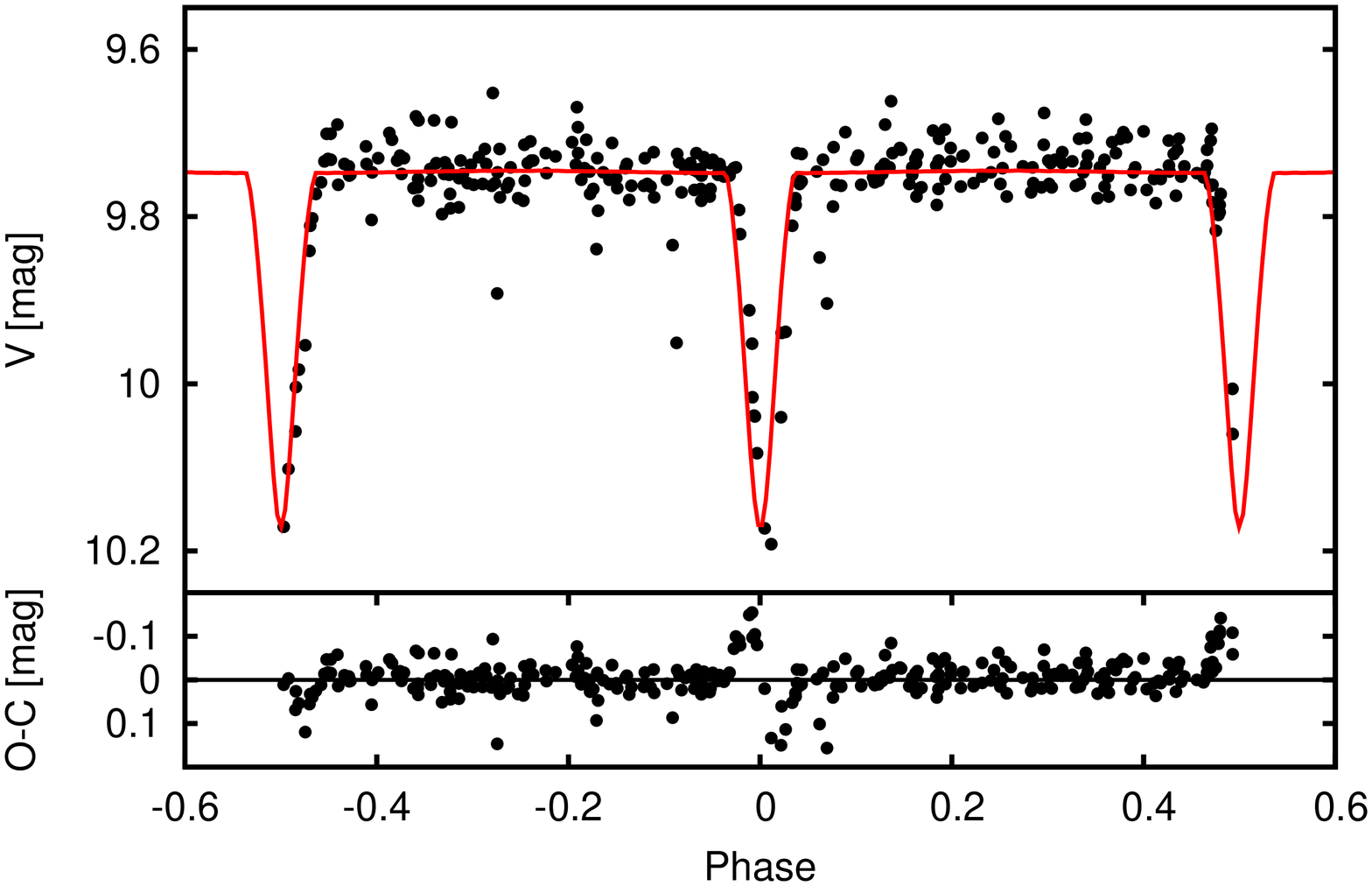}
\includegraphics[angle = -90, bb = 108 63 548 752, width=\columnwidth]{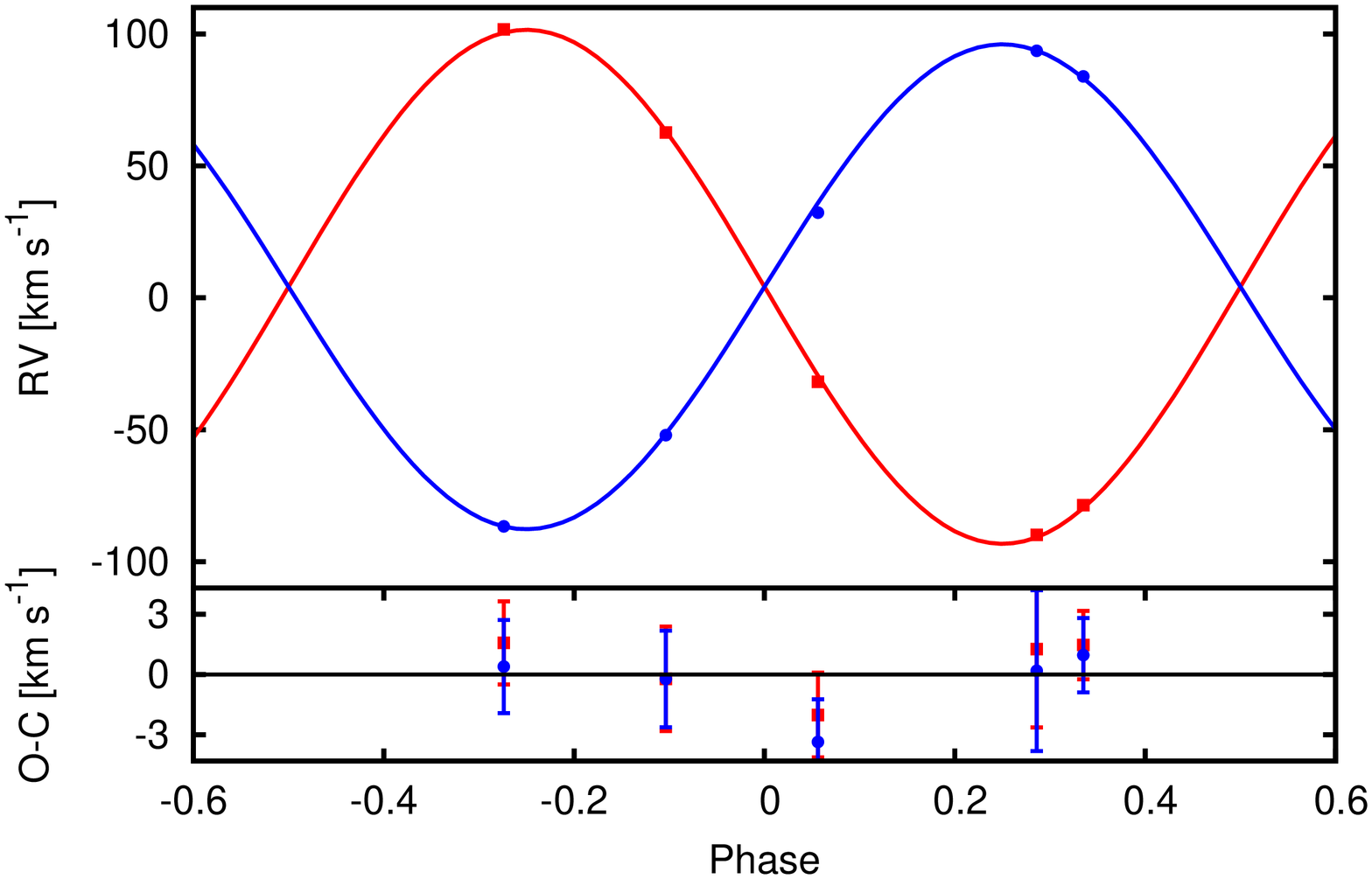}
\caption{PHOEBE model of A023631+1208: ACVS light curve (top panel) and radial velocities obtained with BACHES (bottom panel).}
\label{fig:HD10700-rvmodel}
\end{center}
\end{figure}

A model of A023631+1208.6 has been published by \cite{Helminiak2009}. We followed the same methodology as in the cited paper but we used BACHES spectroscopic data instead of the data from the AAT. In both cases photometry data came from ACVS. We present the resulting radial velocity and light curves in Fig. \ref{fig:HD10700-rvmodel}.  Masses  obtained from the model are $M_1= 1.24\pm0.04$~M$_{\text{sun}}$ and $M_2= 1.31\pm0.04$~M$_{\text{sun}}$ compared to $M_1= 1.357\pm0.008$~M$_{\text{sun}}$ and $M_2= 1.138\pm0.007$~M$_{\text{sun}}$ that have been previously published. The error of mass determination is 3\% for BACHES data and 0.6\% for AAT data. With our relatively inexpensive setup we were able to derive stellar masses with a precision less than one order of magnitude worse than in case of a 3.9-m telescope and a $R\sim60,000$ spectrograph. In both cases Th-Ar spectra were used for wavelength calibration. We have used \textsc{jktebop} and \textsc{jktabsdim} \citep{Southworth2004a}, \textsc{phoebe} \citep{Prsa2005} and our own numerical codes to obtain the model.

\section{Summary}
\label{sec:Summary}
We have been able to obtain spectra with an SNR of ~20 at 5500 \AA\ for a 10th mag star in a 30 minute exposure at a resolution of 20,000. Considering a small, 0.5-m primary mirror and not perfect guiding, this result is very good for a compact slit-based spectrograph. We expect that better guiding and a larger slit (50~$\mu$m instead of 25~$\mu$m) along with a higher QE camera will allow us to reach 11th mag stars with an acceptable SNR. Increasing the width of the slit will reduce the resolution of the spectrograph, but even then the instrument will be adequate for a survey-type project. There are, however, a few points that need to be addressed. The spectroscopic CCD camera mount interface needs to be modified. A dedicated, oversized diameter flange should be screwed directly into BACHES' housing to minimise the flexures. Standard connectors used in amateur astronomy (1.25 and 2.0 inch diameters) do not provide enough support even for medium-sized cooled CCD cameras used for spectroscopy. The FLI filter wheel's mechanical interface needs to be stiffened to reduce the flexure of the filter wheel and CCD camera assembly. The interface should be split into two independent mechanical parts: one that screws into the filter wheel and ends with a 4" male dovetail ASA interface. The GAM should have a corresponding female 4" dovetail interface instead of the FLI thread.  This will allow easy rotation, assembly and disassembly of the imaging train. 

Despite the minor mechanical shortcomings that are visible in the tests, the resulting precision of radial velocity measurements is in order of 1.5 km~s$^{-1}$ which a very good result. The stability of the spectrograph is satisfactory when comparing actual shifts of Th-Ar spectra taken before and after the object's spectrum. \cite{Konacki2003} show shifts of the Th-Ar calibration spectra on the CCD of the Keck I High Resolution Echelle Spectrometer (HIRES) throughout a four day observing run. HIRES is mounted on the Nasmyth platform and an amplitude of 1 pixel corresponding to 24 $\mu$m is visible. In this context the stability of our setup is excellent.

\cite{Helminiak2009} present radial velocities obtained with the Grating Instrument for Radiation Analysis with a Fibre-Fed Echelle (GIRAFFE) for two eclipsing binaries: A010538-8003.7, 10.01 mag and A174626-1153.0, 10.81 mag). The rms of the best fits are 3.02 and 3.11 km~s$^{-1}$ for A010538-8003.7, 1.92 and 1.52 km~s$^{-1}$ for A174626-1153.0, compared with 1.6 and 1.7 km~s$^{-1}$  obtained with BACHES for A023631+1208.6, a 9.75 mag binary. GIRAFFE is a low-cost fibre-fed instrument located in the Coud\'{e} room in a much more stable environment than BACHES.

Our overall impression is that the tested setup is a very capable and can produce science-grade results. The commissioning time is very short, most of the components work out-of-the-box. A fully automated spectroscopic mode will require a significant amount of coding time but is definitely possible to implement without major hardware modifications. The total time on site needed to mount the spectrograph, connect all necessary cables and take stellar spectra is about 3-4 hours which is a very impressive result.

Comparing results obtained with BACHES with other instruments is not straightforward due to the number of parameters that influence the overall performance of the observing systems -- apart from the spectrograph, also the telescopes and cameras have strong impact on the capabilities of the setups. However, some conclusions can be formulated when comparing it with PUCHEROS and eShel. BACHES is the only spectrograph in this trio that is mounted directly on the imaging train of the telescope. This means that its position and orientation change constantly throughout the night. The spectrograph also operates in continuously changing ambient conditions -- temperature and pressure variations influence the mechanical stability of the instrument. On the other hand, BACHES is a slit spectrograph, so light is not lost in the fibre that feeds the other two instruments. As a result, it is possible to obtain a spectrum of a $\sim 10$ mag star with a SNR 20 with 30 minute exposures using BACHES. Based on the information provided in \cite{Vanzi2012} and \cite{Csak2014} we can estimate that PUCHEROS and eShel will require $\sim$ 1 hour exposures to achieve a comparable SNR. It should be noted, however, that simple scaling of the exposure time towards longer integration times is just an optimistic estimate. During long exposures the performance of the telescope's mount plays an important role in the overall efficiency of the system as the tracking errors influence the amount of light that enters the slit/fibre. In Project Solaris we use gear-less direct drive mounts with very high dynamical performance. In terms of stability eShel is roughly one order of magnitude better than BACHES and PUCHEROS for similar targets. Comparison has been made for 0.5-m telescopes.

BACHES exceeded our expectations in terms of its usefulness in radial velocity measurements for binary stars and we plan to equip two Solaris telescopes with these instruments in the near future.

\section*{acknowledgements} 
We would like to thank the CAOS group (Jes\'{u}s Rodr\'{i}guez, Carlos Guirao, Gerardo \'{A}vila, Vadim Burwitz.) and the Baader Planetarium company for allowing us to use a pre-production spectrograph for evaluation. We would like to thank Hector Molina from the CASLEO observatory for making the tests of the spectrograph possible. We would like to thank the staff in  CASLEO, Director Ricardo Gil-Hutton and Ing. Jose-Luis Giuliani in particular, for help during the installation phase of the Solaris-4 observatory.

This work is supported by the European Research Council through a Starting Grant, the National Science Centre through grant 5813/B/H03/2011/40, the Polish Ministry of Science and Higher Education through grant 2072/7.PR/2011/2 and the Foundation for Polish Science through \textit{Idee dla Polski} funding scheme. K.G.H. acknowledges support provided by the National Astronomical Observatory of Japan as Subaru Astronomical Research Fellow, and the Polish National Science Center grant 2011/03/N/ST9/01819. P.S. acknowledges support provided by the National Science Center through grant 2011/03/N/ST9/03192. M.R. acknowledges support provided by the National Science Center through grant 2011/01/N/ST9/02209.

\bibliographystyle{mn2e} 
\bibliography{Baches_MNRAS-ver7astroph}
\label{lastpage}
\end{document}